\documentclass[11pt]{article}

\usepackage{geometry}                
\usepackage{graphicx}
\usepackage{amsmath,amssymb,amsfonts,amsthm,bm,mathtools}
\usepackage{tikz-cd} 
\usepackage{cite}
\usepackage{authblk} 
\usepackage{hyperref} 

\hypersetup{%
  colorlinks=true,
  linkcolor=blue,
  linkbordercolor=red,
  citecolor=red,
}

\newtheorem{theorem}{Theorem}
\newtheorem{prop}{Proposition}

\newtheorem{comm}{Comment}
\newtheorem{lemma}{Lemma}
\newtheorem{ex}{Example}

\title{On finite-dimensional smoothed-particle Hamiltonian reductions of the Vlasov equation}
\author[1]{William Barham}
\author[2]{Philip J. Morrison}
\affil[1]{Oden Institute for Computational Engineering and Sciences, The University of Texas at Austin}
\affil[2]{Department of Physics and Institute for Fusion Studies, The University of Texas at Austin}
\date{\today}                     
\begin{document}

\maketitle

\section*{Abstract}
The inclusion of spatial smoothing in finite-dimensional particle-based Hamiltonian reductions of the Vlasov equation are considered. In the context of the Vlasov-Poisson equation (and other mean-field Lie-Poisson systems), smoothing amounts to a convolutive regularization of the Hamiltonian. This regularization may be interpreted as a change of the inner product structure used to identify the dual space in the Lie-Poisson Hamiltonian formulation. In particular, the shape function used for spatial smoothing may be identified as the kernel function of a reproducing kernel Hilbert space whose inner product is used to define the Lie-Poisson Hamiltonian structure. It is likewise possible to introduce smoothing in the Vlasov-Maxwell system, but in this case the Poisson bracket must be modified rather than the Hamiltonian. The smoothing applied to the Vlasov-Maxwell system is incorporated by inserting smoothing in the map from canonical to kinematic coordinates. In the filtered system, the Lorentz force law and the current, the two terms coupling the Vlasov equation with Maxwell's equations, are spatially smoothed. 

\tableofcontents

\section{Introduction}

When studying the dynamics of infinite-dimensional Hamiltonian systems, it is frequently advantageous to find flows which are described by a finite amount of information. This is the idea behind a finite-dimensional Hamiltonian reduction. A particular kind of Hamiltonian reduction relevant to the study of fluids and plasmas is a particle-based reduction which reduces the model to a system of ordinary differential equations for the particle characteristics. The earliest kind of particle-based reduction of an infinite-dimensional Hamiltonian system (although not recognized as such when first introduced) is the point-vortex method for two-dimensional incompressible flows; see \cite{Kraichnan_1980} and references therein for a brief historical account. The Hamiltonian structure of the two-dimensional vorticity equation is discussed in \cite{pjm82, osti_6351319}, and one may show that its treatment via point-vorticies may be understood as a finite-dimensional Hamiltonian reduction. Moreover, many incompressible two-dimensional fluid models, such as the Charney-Hasegawa-Mima equation \cite{10.1063/1.3194275, CHANDRE2014956}, are likewise amenable to treatment with point-vortex methods. Particle-based Hamiltonian reductions of the Vlasov equation are relevant in computational plasma physics because the particle-in-cell (PIC) method \cite{doi:10.1201/9780367806934, doi:10.1201/9781315275048}, when appropriately formulated, may in certain cases be interpreted as a particle-based reduction of the continuous Hamiltonian structure \cite{kraus_kormann_morrison_sonnendrücker_2017}. Particle-based representations of the two-dimensional vorticity field or of the Vlasov phase-space density make an Ansatz that these fields may be represented as a weighted sum of Dirac delta functions. This paper considers the consequences of spatially smoothing this representation via convolution with a smooth kernel function. While this paper is concerned with the general concept of regularizing particle-based Hamiltonian reductions of the Vlasov equation and related models, the motivation for doing so is primarily as a tool for the analysis and design of smoothed PIC methods. 

While the PIC method is convenient in that it discretizes the high-dimensional phase space distribution function via a particle-based representation while treating the electromagnetic/electrostatic fields on a grid, therefore avoiding the challenges associated with attempting a grid-based discretization of a high-dimensional field and yielding a method highly amenable to parallelization, this Monte-Carlo-type discretization of the phase space distribution function suffers from substantial statistical noise. Many PIC methods therefore employ some form of smoothing. The smoothing is typically simply interpreted as a regularization step when distributing the particle information to the grid to compute the electrostatic potential from the charge density. A recent development in the design of PIC methods are structure-preserving PIC methods which, in some manner, retain Hamiltonian or variational structure of the continuous problem, e.g. \cite{10.1063/1.4742985, EVSTATIEV2013376, kraus_kormann_morrison_sonnendrücker_2017}. See \cite{10.1063/1.4982054} for a review of structure-preserving methods in plasma physics. If spatial smoothing is applied to a structure-preserving PIC method without considering its impact on the Hamiltonian structure, it is likely that the method no longer be structure-preserving. This work seeks to clarify structure-preserving methods of adding spatial smoothing to PIC methods. 

Smoothed particle reductions of the Vlasov-Poisson system amount to introducing a small-scale regularization of the Hamiltonian. The typical formulation of smoothed PIC methods for the Vlasov-Poisson equation may be shown to correspond with an exact reduction not of the Vlasov-Poisson equation, but rather of a Hamiltonian system in which the Hamiltonian has been regularized by convolving the phase-space distribution with a smoothing kernel while the Poisson bracket remains the same. This regularized model moreover may be equivalently written as a Lie-Poisson Hamiltonian system built from the inner product of a reproducing kernel Hilbert space (RKHS) associated with the symmetric positive definite smoothing kernel. This strategy generalizes to a broad class of mean-field Hamiltonian models \cite{Morrison2003} suggesting smoothed particle reductions for a broad class of models of fluids and plasma. 

The Hamiltonian structure of the Vlasov-Maxwell equation may likewise be spatially smoothed such that its particle-based Hamiltonian reduction is regularized. However, the approach is quite different to that which was used for the Vlasov-Poisson equation. Whereas the Vlasov-Poisson equation adds smoothing by regularizing the Hamiltonian, the Vlasov-Maxwell equation is regularized by adding smoothing to the Poisson bracket. This bracket satisfies the Jacobi identity as it may be derived via a coordinate change from a bracket in canonical form. The filtered Vlasov-Maxwell Hamiltonian structure may be used to derive a PIC discretization based on the Geometric ElectroMagnetic PIC (GEMPIC) method \cite{kraus_kormann_morrison_sonnendrücker_2017} which is an exact finite dimensional Hamiltonian reduction and that this PIC method is identical to one which was formerly derived using a discretized action principle formulation of the Vlasov-Maxwell system \cite{variational_PIC_campos-pinto}.  

\section{On particle-based Hamiltonian reductions}
As this work discusses smoothed particle-based Hamiltonian reductions of certain Hamiltonian field theories from the physics of fluids and plasmas, it is helpful to first introduce the properties of a Hamiltonian system, the definition of a Hamiltonian reduction, and then to provide a brief pedagogical illustration of the notion of a finite-dimensional particle-based Hamiltonian reduction of two-dimensional incompressible fluids. 

Hamiltonian systems are described by:
\begin{itemize}
    \item a manifold $\mathcal{M}$
    \item a Poisson bracket, $\{ \cdot, \cdot \}: C^\infty(\mathcal{M}) \times C^\infty(\mathcal{M}) \to C^\infty(\mathcal{M})$
    \item an energy functional, the Hamiltonian, $H: C^\infty(\mathcal{M}) \to \mathbb{R}$.
\end{itemize}
The evolution for any $F \in C^\infty(\mathcal{M})$ is given by
\begin{equation}
    \dot{F} = \{ F, H \} \,.
\end{equation}
Given another Poisson manifold, $(M, [\cdot, \cdot])$, a Hamiltonian reduction is prescribed by a map $\Pi: \mathcal{M} \to M$ such that
\begin{equation}
    [ F \circ \Pi, G \circ \Pi ] = \{F, G\} \circ \Pi \,.
\end{equation}
If $\text{dim}(\mathcal{M}) = \infty$ and $\text{dim}(M) < \infty$, this is a finite-dimensional Hamiltonian reduction. 

Two-dimensional incompressible fluids enjoy a Hamiltonian formulation \cite{pjm82, osti_6351319}. Let $\Omega \subset \mathbb{R}^2$. A two-dimensional incompressible fluid may be described as follows:
\begin{equation}
    \partial_t \omega + \bm{v} \cdot \nabla \omega = 0 \,, 
    \quad \text{where} \quad
    \omega = \Delta \phi = \partial_x^2 \phi + \partial_y^2 \phi \,,
    \quad \text{and} \quad
    \bm{v} = \nabla^\perp \phi = (- \partial_y \phi, \partial_x \phi) \,.
\end{equation}
This is Hamiltonian with
\begin{equation} \label{eq:vort_ham}
    H[\omega] 
    = - \frac{1}{2} \int_\Omega \omega \phi \mathsf{d}^2 \bm{x} 
    = - \frac{1}{2} \int_\Omega \omega \Delta^{-1} \omega \mathsf{d}^2 \bm{x} \,,
    \quad \text{and} \quad
    \{F, G\} = \int_\Omega \omega \left[ \frac{\delta F}{\delta \omega}, \frac{\delta G}{\delta \omega} \right] \mathsf{d}^2 \bm{x} \,.
\end{equation}
This model admits an exact finite-dimensional Hamiltonian reduction via point-vorticies. Suppose the vorticity field may be represented as
\begin{equation}
    \omega(x, y) = \sum_{a=1}^{N_p} \Gamma_a \delta( x_a - x ) \delta( y_a - y ) \,.
\end{equation}
Plugging in this Ansatz yields a Hamiltonian and Poisson bracket describing the evolution of the centers of each point vortex:
\begin{equation}
    [f, g] = \sum_{a=1}^{N_p} \Gamma_a^{-1} \left( \frac{\partial f}{\partial x_a} \frac{\partial g}{\partial y_a} - \frac{\partial g}{\partial x_a} \frac{\partial f}{\partial y_a} \right) \,,
\end{equation}
and
\begin{equation}
    H \left( \{ x_\alpha, y_\alpha \}_{\alpha = 1}^{N_p} \right) 
    = \frac{1}{4 \pi} \sum_{\alpha=1}^{N_p} 
    \sum_{ \substack{\beta = 1 \\ \beta > \alpha}}^{N_p} \Gamma_\alpha \Gamma_\beta 
    \log \left( (x_\alpha - x_\beta)^2 + (y_\alpha - y_\beta)^2 \right) \,.
\end{equation}
This Hamiltonian arises from the Green's function for the Laplacian in two dimensions because one must solve $\Delta^{-1} \omega$ to compute the energy. A spatially smoothed particle-based reduction would regularize $\omega$ prior to inverting the Laplacian.

\section{The Hamiltonian structure of the Vlasov-Poisson equation}
The Vlasov-Poisson equation has a remarkably similar Hamiltonian structure to two-dimensional vorticity. For this reason, this model also admits particle-based Hamiltonian reductions. Its Lie-Poisson Hamiltonian structure, described in \cite{MORRISON1980383}, is as follows:
\begin{multline}
    \{F, G\} = \int f \left[ \frac{\delta F}{\delta f}, 
        \frac{\delta G}{\delta f} \right] \mathsf{d} x \mathsf{d} v \,, \\
    \quad \text{and} \quad
    H[f] = \frac{1}{2} \int | v |^2 f(x, v) 
        \mathsf{d} x \mathsf{d} v
        +
        \frac{1}{2} \int \phi(x) f(x,v) \mathsf{d} x \mathsf{d} v \,,
\end{multline}
where $- \Delta_x \phi(x) = \int f(x,v) \mathsf{d} v$, and the finite dimensional Poisson bracket is defined 
\begin{equation}
    [f, g] = \nabla_x f \cdot \nabla_v g 
        - \nabla_x g \cdot \nabla_v f \,.
\end{equation}
Being a Lie-Poisson system, one may write this more abstractly in terms of the Lie algebra structure $\mathfrak{g}$. Let $\mathfrak{g}^*$ be the dual space to $\mathfrak{g}$ and denote the abstract duality pairing by $\langle \cdot, \cdot \rangle: \mathfrak{g}^* \times \mathfrak{g} \to \mathbb{R}$. Then assuming the phase space distribution function is a member of the dual space, $f \in \mathfrak{g}^*$, one may write
\begin{equation}
    \{F, G\} = \left\langle f, 
        \left[ \frac{\delta F}{\delta f}, 
        \frac{\delta G}{\delta f} \right] 
        \right\rangle
    \quad \text{and} \quad
    H[f] = \left\langle f, \mathcal{V}(z, f) \right\rangle
\end{equation}
where $\mathcal{V}: \mathbb{R}^6 \times \mathfrak{g}^* \to \mathfrak{g}$ is a primative of the pointwise energy function, $E: \mathbb{R}^6 \times \mathfrak{g}^* \to \mathfrak{g}$, such that
\begin{equation}
    \frac{\delta H}{\delta f} = E(z, f) \,.
\end{equation}
The Hamiltonian for the Vlasov equation is obtained by taking the moment of this primitive energy function. A detailed description of the Lie algebra structure may be found in \cite{MARSDEN1982394}. One generally takes $\mathfrak{g}$ to be contained in some Hilbert space, e.g. $\mathfrak{g} \subset L^2$, and identify $\mathfrak{g}^* \simeq \mathfrak{g}$ through the Riesz representation theorem replacing $\langle \cdot, \cdot \rangle$ with $(\cdot, \cdot): \mathfrak{g} \times \mathfrak{g} \to \mathbb{R}$, the inner product structure on $\mathfrak{g}$. In the case of the Vlasov-Poisson system, the inner product is taken to be the $L^2$ inner product and the pointwise energy is given by
\begin{equation}
    E(z, f) = \frac{1}{2} | v |^2 
        - \Delta_x^{-1} \int f \mathsf{d} v \,.
\end{equation}
This work considers the consequences of letting the inner product structure be that associated with some RKHS, $\mathcal{H}$ (a brief summary of the essential properties of RKHS will be discussed subsequently). That is, take $f \in \mathcal{H}$, and
\begin{equation}
    \{F, G\} = \left( f, 
        \left[ \frac{\delta F}{\delta f}, 
        \frac{\delta G}{\delta f} \right] 
        \right)_{\mathcal{H}}
    \quad \text{and} \quad
    H[f] = \left( f, \mathcal{V}(z, f) \right)_{\mathcal{H}} \,,
\end{equation}
where the functional derivatives in this formulation are taken to mean
\begin{equation}
    \left\langle DF[f], \delta f \right\rangle 
    = \left( \frac{\delta F}{\delta f}, \delta f \right)_{\mathcal{H}} \,.
\end{equation}
This is beneficial because one may take $f \in \mathcal{H}$ to be a sum of delta functions in the RKHS which are precisely the reproducing kernel functions. Hence, this provides a convenient framework for interpreting smoothed PIC methods in a Hamiltonian context. 

\section{Properties of the kernel smoothing operator}
The smoothing of a distribution function via convolution with a kernel operator may be thought of as a coordinate change provided the smoothing kernel possess appropriate properties. Consider the symmetric Fredholm operator $\mathcal{L}: L^2 \to R(\mathcal{L})$ defined by convolution with a symmetric kernel function $K(z | z')$:
\begin{equation}
	\bar{f} \coloneq \mathcal{L} f = \int K(z | z') f(z') \mathsf{d} z' = K_z * f \,,
\end{equation}
where $K_z = K(z | \cdot)$. Then $\mathcal{L}^{-1}: R(\mathcal{L}) \to L^2$ exists and $\mathcal{L}^\dagger = \mathcal{L}$. Note, one cannot in general write $\mathcal{L}^{-1}$ explicitly as it is the solution of an integral equation, but it frequently takes the form of a pseudodifferential operator (e.g. if $K$ is the Green's function for that operator). Being the range of a convolution operator, $R(\mathcal{L})$ is a space of functions with the regularity of $K$. Further, there is a simple criterion which guarantees that $R(\mathcal{L})$ consist of square integrable functions. 
\begin{prop}
Let $k \in L^1$, and suppose the kernel function is isotropic: $K(z|z') = k( z - z' )$. Then $R(\mathcal{L}) \subset L^2$. 
\end{prop}
\noindent \textit{Proof:} This is a simple consequence of Young's inequality:
\begin{equation}
	\| \bar{f} \|_{L^2} = \| k * f \|_{L^2} \leq \| k \|_{L^1} \| f \|_{L^2}. 
\end{equation}
\qed

\noindent It shall be assumed in general that $k \in L^1$ and $K(z|z') = k( z - z' )$. Further, it is convenient to let $\int k \mathsf{d} z = 1$ since it will be necessary that the kernel function in some manner approximate the delta function. It is not necessary to insist that the kernel be positive, although this will frequently be the case. 

Under appropriate conditions, one may equip $\mathcal{H} \coloneq R(\mathcal{L})$ with a Hilbert space structure. To this end, assume that $\mathcal{L}^{-1}$ is positive definite. Define an inner product structure $(\cdot, \cdot)_{\mathcal{H}}: \mathcal{H} \times \mathcal{H} \to \mathbb{R}$ such that
\begin{equation}
	(f,g)_{\mathcal{H}} = (f, \mathcal{L}^{-1} g)_{L^2}. 
\end{equation}
Since, as shown above, $\mathcal{H} \subset L^2$, this is well defined. That this is an inner product follows from the assumption that $\mathcal{L}^{-1}$ is positive definite. One should further demonstrate completeness of this Hilbert space. However, it is easier to differ this consideration until the following section in which this inner product space is identified with the RKHS associated with the smoothing kernel. 

\begin{ex}
To make this discussion concrete, consider the following example in one spatial dimension. Suppose the smoothing kernel is a Laplace distribution:
\begin{equation}
	K(x | x') = k(x - x') = \frac{1}{2 \alpha} \exp \left( - \frac{ | x - x' | }{\alpha } \right) \,,
\end{equation}
for $\alpha > 0$. One may show that $\int_{\mathbb{R}} k(x) \mathsf{d} x = 1$. Letting $\mathcal{L}f = k * f$, one may show $\mathcal{L}^{-1} = 1 - \alpha^2 \partial_x^2$. Hence, in this case, the Hilbert space is a modified $H^1$ Sobolev space with inner product
\begin{equation}
	(f, g)_{\mathcal{H}} = (f,g)_{L^2} + \alpha^2 (\partial_x f, \partial_x g)_{L^2} \,.
\end{equation} 
\end{ex}

While it is possible to build the theory entirely in terms of the classical theory of integral operators, one may leverage the theory of RKHS to show that any symmetric positive definite smoothing kernel may be used. Given any symmetric positive definite kernel $K$, the Moore-Aronszajn theorem \cite{aronszajn1950} implies there exists an RKHS, denoted $(\mathcal{H}, (\cdot, \cdot)_{\mathcal{H}}$), for which $K$ is the reproducing kernel. That is, $\forall f \in \mathcal{H}$, $f(z) = (K_z, f)_{\mathcal{H}}$ where $K_z = K(z | \cdot)$. In other words, $K_z \in \mathcal{H}$ acts as the delta function in the RKHS. What makes RKHS special is that the delta function is a member of the Hilbert space itself; contrast this with $L^2$ where the Dirac delta function is not a member of the Hilbert space but rather a functional whose domain is a strict subset of the Hilbert space. Stating the theory using the language of RKHS simplifies the discussion and notation somewhat and, as will be seen, the fact that convolution with the kernel function is the pointwise evaluation operator in the RKHS provides a convenient interpretation of taking moments of the phase space distribution in a filtered PIC method. 

One may connect the RKHS perspective with the previous formulation in terms of integral operators by specializing to kernels of the form $K(z | z') = k(z - z')$. Recalling the definition $\mathcal{L}f = k * f$ and defining
\begin{equation}
	(f, g) \coloneq (f, \mathcal{L}^{-1} g)_{L^2} \,,
\end{equation}
one finds that
\begin{equation}
	(K_z, f) = (K_z, \mathcal{L}^{-1} f)_{L^2} = \mathcal{L} \mathcal{L}^{-1} f = f \,.
\end{equation}
By the uniqueness of the RKHS associated with a given kernel, it follows that $(\cdot, \cdot) \equiv (\cdot, \cdot)_{\mathcal{H}}$. Hence, one need not be concerned with the precise form of $\mathcal{L}^{-1}$, or the inner product which is weighted by inverse integral operator. It is enough to know that there does exist an appropriate Hilbert space for any symmetric positive definite kernel one chooses. 

\section{A filtered particle-based reduction of the Vlasov-Poisson bracket}
A Vlasov-Poisson PIC-method which uses spatial smoothing may be thought of as an exact reduction of the continuous Poisson bracket in an appropriate Hilbert space. Define $\bar{f}$ where
\begin{equation}
    \bar{f} = \mathcal{L} f = k * f \,,
    \quad \text{where} \quad
    f \in L^2 \,.
\end{equation}
The ``bar" notation will be consistently used to denote distribution functions in the filtered space, $\mathcal{H}$. To begin, it is necessary to give several tools to facilitated functional calculus in the RKHS.
\begin{lemma}
Let $(\mathcal{H}, (\cdot, \cdot)_{\mathcal{H}})$ be the RKHS associated with the symmetric positive definite kernel $K$. Define the filtered particle representation of a distribution function by
\begin{equation}
	\bar{f}(z) = \sum_{a=1}^{N_p} w_a K(z | z_a ) \,.
\end{equation}
Denote $Z = (z_1, z_2, \hdots, z_{N_p})$. Then
\begin{equation}
	(\bar{f}, g)_{\mathcal{H}} = \sum_{a = 1}^{N_p} w_a g(z_a) \,.
\end{equation}
Further, the first variation of $\bar{f}$ with respect to $Z$, $D\bar{f}(Z) \cdot \delta Z \in \mathcal{H}$, acts on $g \in \mathcal{H}$ according to
\begin{equation}
	\left( D\bar{f}[Z] \cdot \delta Z, g \right)_{\mathcal{H}} = - \sum_a w_a \nabla_z g(z_a) \cdot \delta z_a \,.
\end{equation}
\end{lemma}

\noindent \textit{Proof:} That 
\begin{equation}
	(\bar{f}, g)_{\mathcal{H}} = \sum_{a = 1}^{N_p} w_a g(z_a) \,
\end{equation}
follows from the fact that the inner product is bilinear and $K_{z_a}$ is the evaluation functional. The weak derivative of $K_{z_a}$ is obtained by integrating by parts:
\begin{multline}
	\left| \left( K_{z_a + \delta z_a} - K_{z_a} - \nabla_z K_{z_a} \cdot \delta z_a, g \right)_{\mathcal{H}} \right|
		= \left| \left( K_{z_a + \delta z_a} - K_{z_a}, g \right)_{\mathcal{H}} + \left( K_{z_a}, \nabla_z g \cdot \delta z_a \right)_{\mathcal{H}} \right| \\
		= \left| g(z_a + \delta z_a) - g(z_a) + \nabla_z g(z_a) \cdot \delta z_\alpha \right| 
		= O(| \delta z_a | ) \quad \forall g \in \mathcal{H} \,,
\end{multline}
where $g$ is assumed to be differentiable. The result follows from the linearity of the derivative operator. \qed

\begin{comm}
One can see that taking moments with respect to the phase space distribution using the RKHS inner product simply corresponds to pointwise evaluation. Moreover, derivatives of this evaluation operator in an RKHS behave exactly as derivatives of the Dirac delta function: $( \delta'(x-x_a), g)_{L^2} = -g'(x_a)$.
\end{comm}

Given $\bar{F}: \mathcal{H} \to \mathbb{R}$, the functional derivative $D\bar{F}[\bar{f}]: \mathcal{H} \to \mathbb{R}$ is defined
\begin{equation}
	\| \bar{F}[\bar{f} + \delta \bar{f}] - \bar{F}[\bar{f}] - D\bar{F}[\bar{f}] \delta \bar{f} \|_{\mathcal{H}} = \| \delta \bar{f} \|_{\mathcal{H}} \,,
\end{equation}
where $\delta \bar{f} \in \mathcal{H}$ is arbitrary and $\| \cdot \|_{\mathcal{H}}$ is the Hilbert space norm. One may in turn identify the functional derivative operator with an element of $\mathcal{H}$ using the Riesz representation theorem:
\begin{equation}
	D\bar{F}[\bar{f}] \delta \bar{f} = \left( \frac{\delta^{\mathcal{H}} \bar{F}}{\delta \bar{f}}, \delta \bar{f} \right)_{\mathcal{H}} \,.
\end{equation}
Hence, $\delta^{\mathcal{H}} \bar{F} / \delta \bar{f} \in \mathcal{H}$. 

\begin{comm}
The notation $\delta^{\mathcal{H}} \bar{F} / \delta \bar{f}$ is used to refer to the identification of the functional derivative using the $\mathcal{H}$ inner product whereas $\delta F/\delta f$ will denote the more usual identification with the $L^2$ inner product. 
\end{comm}

\noindent The $\mathcal{H}$ functional derivative may be related to the standard $L^2$ functional derivative as follows. Let $\bar{F}[\bar{f}] = F[f]$. Then since $\bar{f} = \mathcal{L} f$, one finds
\begin{equation}
	D\bar{F}[\bar{f}] \delta \bar{f} 
		= \left( \frac{\delta^{\mathcal{H}} \bar{F}}{\delta \bar{f}}, \delta \bar{f} \right)_{\mathcal{H}} 
		= \left( \frac{\delta^{\mathcal{H}} \bar{F}}{\delta \bar{f}}, \mathcal{L}^{-1} \delta \bar{f} \right)_{L^2} 
		= \left( \frac{\delta F}{\delta f}, \delta f \right)_{L^2} \, 
		= DF[f] \delta f \,.
\end{equation}
Hence, $\delta^{\mathcal{H}} \bar{F} / \delta \bar{f} \in \mathcal{H}$, and it is related to the standard $L^2$ functional derivative via
\begin{equation}
	\frac{\delta^{\mathcal{H}} \bar{F}}{\delta \bar{f}} = \mathcal{L} \frac{\delta F}{\delta \bar{f}}
	\quad \text{and} \quad \frac{\delta^{\mathcal{H}} \bar{F}}{\delta \bar{f}} = \frac{\delta F}{\delta f} \,. 
\end{equation}

These results may be used to rewrite the Vlasov-Poisson bracket in terms of the RKHS inner product structure. To begin, recall the standard formulation of the Vlasov-Poisson bracket:
\begin{equation}
	\{F, G\} = \int f \left[ \frac{\delta F}{\delta f}, \frac{\delta G}{\delta f} \right] \mathsf{d} z = \left( f, \left[ \frac{\delta F}{\delta f}, \frac{\delta G}{\delta f} \right] \right)_{L^2}
\end{equation}
where $[\cdot, \cdot]$ is the finite dimensional canonical bracket. From this, one immediately obtains the following result.
\begin{prop}
Let $\bar{f} = \mathcal{L} f$. Then
\begin{equation}
	\{F, G\} 
		= \left( f, \left[ \frac{\delta F}{\delta f}, \frac{\delta G}{\delta f} \right] \right)_{L^2}
		= \left( \bar{f}, \left[ \frac{\delta^{\mathcal{H}} \bar{F}}{\delta \bar{f}}, \frac{\delta^{\mathcal{H}} \bar{G}}{\delta \bar{f}} \right] \right)_{\mathcal{H}} 
		\eqcolon \{ \bar{F}, \bar{G} \}_{\mathcal{H}} \,.
\end{equation}
\end{prop}

\noindent As the two different representations are really only a change of variables, the Jacobi identity is satisfied when the bracket is written with respect to the RKHS inner product structure. 

Consider finite dimensional reductions of the Poisson bracket using a smoothed particle representation.
\begin{lemma}
Let $\bar{f}$ be the filtered-PIC representation as defined in the previous lemma, and let $\bar{F}: \mathcal{H} \to \mathbb{R}$. Moreover, let $\bar{F}[\bar{f}] = F_p(Z)$. Then
\begin{equation}
	\frac{\partial F_p}{\partial z_a} = - w_a \nabla_z \left. \frac{\delta^{\mathcal{H}} \bar{F}}{\delta \bar{f}} \right|_{z_a} \,.
\end{equation}
\end{lemma}
\noindent \textit{Proof:} By definition, one has
\begin{equation}
	D\bar{F}[ \bar{f} ] \delta \bar{f} = \left( \frac{\delta^{\mathcal{H}} \bar{F}}{\delta \bar{f}}, \delta \bar{f} \right)_{\mathcal{H}} = \sum_a \frac{\partial F_p}{\partial z_a} \cdot \delta z_a \,.
\end{equation}
The chain rule implies
\begin{equation}
	\left( \frac{\delta \bar{F}}{\delta \bar{f}}, \delta \bar{f} \right)_{\mathcal{H}}
		= \sum_a \left( \frac{\delta \bar{F}}{\delta \bar{f}}, D \bar{f} \cdot \delta z_a \right)_{\mathcal{H}}
		= - \sum_a \left( w_a \left. \nabla_z \frac{\delta \bar{F}}{\delta \bar{f}} \right|_{z_a} \cdot \delta z_a \right).
\end{equation}
It follows that
\begin{equation}
	\begin{aligned}
		\sum_a \left[ \frac{\partial F_p}{\partial z_a} \cdot \delta z_a \right]
			= -\sum_a \left[ w_a \nabla_z  \left. \frac{\delta \bar{F}}{\delta \bar{f}} \right|_{z_a} \cdot \delta z_a \right]
		\implies
			\frac{\partial F_p}{\partial z_a} = - w_a \nabla_z \left. \frac{\delta \bar{F}}{\delta \bar{f}} \right|_{z_a} \,.
	\end{aligned}
\end{equation}
\qed

\noindent From this, one immediately has the following result.
\begin{prop}
Let $\{\bar{F}, \bar{G}\}_{\mathcal{H}}$ be the Vlasov-Poisson bracket as expressed in $\mathcal{H}$, and let $\bar{f} = \sum_a w_a K_{z_a}(z)$. Then
\begin{equation}
	\{\bar{F}, \bar{G}\}_{\mathcal{H}} 
		= \left( \bar{f}, \left[ \frac{\delta^{\mathcal{H}} \bar{F}}{\delta \bar{f}}, 
			\frac{\delta^{\mathcal{H}} \bar{G}}{\delta \bar{f}} \right] \right)_{\mathcal{H}}
		= \sum_a \frac{1}{w_a} \left[ \frac{\partial F_p }{\partial Z}, \frac{\partial G_p }{\partial Z} \right]
\end{equation}
where $\partial F_p/\partial Z = \left( \partial F_p/\partial z_1, \hdots, \partial F_p/\partial z_{N_p} \right)$. 
\end{prop}

The results in this section should not be surprising as it was previously known that a representation of the distribution function in terms of delta functions yields an exact reduction of the Vlasov-Poisson bracket. One has merely changed coordinates via a convolution integral operator. What is gained from this effort will be seen in the next section which shows that filtered-PIC methods are exact reductions of a modified Vlasov-Poisson equation. 

\section{A Hamiltonian filtered-PIC formulation for the Vlasov-Poisson equation}
This section considers a generic filtered-PIC method for the Vlasov-Poisson equation and show that this may be thought of as an exact reduction of a suitably modified continuous Vlasov-Poisson system. For convenience, all physical constants have been set equal to unity. Consider the particle discretization
\begin{equation}
\begin{cases}
	\dot{x}_a = v_a \\
	\dot{v}_a = \nabla_{x_a} \phi (x_a) \,,
\end{cases}
\quad \forall a \in \{1, \hdots, N_p \} \,,
\end{equation}
where $ - \Delta_{x_a} \phi (x_a) = \sum_{b \neq a} w_b K_{x_a}(x_b)$ and $K_{x_a}$ is a smoothing kernel for the spatial portion of the distribution function. Notice that the Poisson equation has not been discretized. One may use a finite difference or finite element discretization to do so. It is more convenient to leave the method employed in the Poisson solver unspecified as the precise method used is not important for the purposes of this discussion. 

The shape function for the distribution function employed by this particle method is
\begin{equation}
	\bar{f}(x,v) = \sum_a w_a K(x | x_a) \delta(v - v_a) \,,
\end{equation}
and the RKHS is one in which the spatial dimensions of the $L^2$ inner product have been weighted by the inverse convolution operator, but not the velocity dimensions. This is done because it is conventional to smooth in the spatial dimensions only, but there is nothing to prevent smoothing over all of phase-space. It would be worth studying smoothing in velocity space in a future work. That is,
\begin{equation}
	(\bar{f}, \bar{g})_{\mathcal{H}}
		= \int \bar{f} \mathcal{L}^{-1} \bar{g} \mathsf{d} x \mathsf{d} v
\end{equation}
where $\mathcal{L} f = \int K(x| x') \delta(v - v') f(x',v') \mathsf{d} x' \mathsf{d} v'$. First, a Hamiltonian formulation for a regularized Vlasov-Poisson system in the RKHS is derived, and then this model is subsequently shown to reduce to the above filtered-PIC method.  

The Hamiltonian for the continuous Vlasov-Poisson system is
\begin{equation}
	H[f] = \frac{1}{2} \int ( | v |^2 + \phi ) f \mathsf{d} x \mathsf{d} v \,,
	\quad \text{where} \quad
	- \Delta_x \phi = \int f \mathsf{d} v \,.
\end{equation}
It is convenient to rewrite the potential energy as follows:
\begin{equation}
\begin{aligned}
	\frac{1}{2} \int \phi f \mathsf{d} z 
		&= - \frac{1}{2} \int \Delta_x^{-1} f(x,v') f(x, v) \mathsf{d} v' \mathsf{d} x \mathsf{d} v \\
		&= - \frac{1}{2} \left( \int f \mathsf{d} v, \Delta_x^{-1} \int f \mathsf{d} v \right)_{L^2_x} \,,
\end{aligned}
\end{equation}
where $L^2_x$ is the spatial $L^2$ inner product. One may abstractly write the Hamiltonian as
\begin{equation}
    H[f] = \left( K(z) + \mathcal{O}(z) f, f \right)_{L^2} \,,
\end{equation}
where the pointwise kinetic energy is given by $K(z) = \frac{1}{2} | v |^2$, and the potential energy is a quadratic form with the self-adjoint operator $\mathcal{O}(z) f = \frac{1}{2} \Delta_{x}^{-1} \int f \mathsf{d} v$. The pointwise energy is $E(z, f) = \delta H/\delta f = K(z) + \Delta_{x}^{-1} \int f \mathsf{d} v$. Hence, one can see that the energy is obtained by taking a moment of a pointwise energy primitive with respect to the distribution function. 

If one were to make a proper change of coordinates $f \mapsto \bar{f}$, one would find that the potential energy becomes
\begin{equation}
	- \frac{1}{2} \left( \int f \mathsf{d} v, \Delta_x^{-1} \int f \mathsf{d} v \right)_{L^2_x}
	\mapsto
	- \frac{1}{2} \left( \int \mathcal{L}^{-1} \bar{f} \mathsf{d} v, \Delta_x^{-1} \int \mathcal{L}^{-1} \bar{f} \mathsf{d} v \right)_{L^2_x} \,.
\end{equation}
However, to recover the above filtered-PIC method, one instead redefines the filtered potential energy to be
\begin{equation}
	- \frac{1}{2} \left( \int \mathcal{L}^{-1} \bar{f} \mathsf{d} v, \Delta_x^{-1} \int \bar{f} \mathsf{d} v \right)_{L^2_x}
		= - \frac{1}{2} \left( \bar{f}, \Delta_x^{-1} \int \bar{f} \mathsf{d} v \right)_{\mathcal{H}} \,.
\end{equation}
Hence, the filtered Hamiltonian is defined to be
\begin{equation}
\begin{aligned}
    \bar{H}_F [\bar{f}] 
    &= \left( \frac{1}{2} \left[ | v |^2 
    - \Delta_x^{-1} \int \bar{f} \mathsf{d} v \right],
    \bar{f} \right)_{\mathcal{H}} 
    = \left( K(z) + \mathcal{O}(z) \bar{f}, 
    \bar{f} \right)_{\mathcal{H}} \\
    &= \frac{1}{2} \int \left( | v |^2 - \Delta_x^{-1} \int \bar{f}(x, v') \mathsf{d} v' \right) \mathcal{L}^{-1} \bar{f}(x, v) \mathsf{d} x \mathsf{d} v \,.
\end{aligned}
\end{equation}
Notice that, $\bar{H}_F [\bar{f}] \neq H[f]$ despite the formal similarity of the two expressions:
\begin{equation}
    H[f] = \left( K(z) + \mathcal{O}(z) f, f \right)_{L^2} \,,
    \quad \text{and} \quad
    \bar{H}_F [\bar{f}] 
        = \left( K(z) + \mathcal{O}(z) \bar{f}, 
            \bar{f} \right)_{\mathcal{H}} \,.
\end{equation}
One has made the formal change of rewriting the Hamiltonian functional by changing the inner product structure while keeping the pointwise energy function the same, but this is not a simple coordinate change. Assuming the smoothing kernel is an approximation of the Dirac delta, this modification is a regularization at small scales. Hence, Hamiltonian structure of the modified Vlasov equation in the RKHS may be written
\begin{equation}
    \{\bar{F}, \bar{G}\}_{\mathcal{H}} 
		= \left( \bar{f}, \left[ \frac{\delta^{\mathcal{H}} \bar{F}}{\delta \bar{f}}, 
			\frac{\delta^{\mathcal{H}} \bar{G}}{\delta \bar{f}} \right] \right)_{\mathcal{H}}
   \quad \text{and} \quad
   \bar{H}_F [\bar{f}] 
        = \left( K(z) + \mathcal{O}(z) \bar{f}, 
            \bar{f} \right)_{\mathcal{H}} \,.
\end{equation}
The equations of motion are then obtained in the usual manner: i.e. for any functional $\bar{F}$ of the filtered phase space distribution, $\bar{f}$, its evolution is given by $\dot{\bar{F}} = \{\bar{F}, \bar{H}_F \}_{\mathcal{H}}$.

It is necessary that the Laplacian be self-adjoint with respect to the RKHS inner product for the potential energy to behave analogously in the RKHS to how it does in the standard $L^2$ space. The following lemma provides conditions which guarantee this.
\begin{lemma} \label{lemma:self_adjoint}
The inverse Laplacian is self-adjoint with respect to this inner product if the spatial smoothing kernel is isotropic and symmetric, i.e. if $K(x, x') = k( | x - x' | )$. 
\end{lemma}
\noindent \textit{Proof:} If the kernel is isotropic one has
\begin{equation}
	K(x,x') = k( | x - x' | ) \implies \Delta_x^{-1} k( | x - x' | ) = \Delta_{x'}^{-1} k( | x - x' | ) \,.
\end{equation}
Now, consider $\bar{f}, \bar{g} \in \mathcal{H}$, and let $\bar{f} = \mathcal{L} f$ and $\bar{g} = \mathcal{L} g$. Then
\begin{equation}
\begin{aligned}
	(\bar{f}, \Delta_x^{-1} \bar{g})_{\mathcal{H}} 
		&= ( \mathcal{L} f, \Delta_x^{-1} \mathcal{L} g)_{\mathcal{H}}
		= ( f, \Delta_x^{-1} \mathcal{L} g )_{L^2} \\
		&= \int f(x, v) \Delta_x^{-1} k( | x - x' | ) \delta( v - v' ) g(x', v') \mathsf{d} x \mathsf{d} v \mathsf{d} x' \mathsf{d} v' \\
		&= \int \Delta_{x'}^{-1} k( | x' - x | ) \delta( v' - v ) f(x, v) g(x', v') \mathsf{d} x \mathsf{d} v \mathsf{d} x' \mathsf{d} v' \\
		&= ( \Delta_x^{-1} \mathcal{L} f, g )_{L^2} 
		= ( \Delta_x^{-1} \bar{f}, \bar{g} )_{\mathcal{H}} \,.
\end{aligned}
\end{equation}
\qed

\noindent As the kernel was already assumed to be symmetric and isotropic, this does not impose any further restrictions on the choice of kernel. The following result immediately follows:
\begin{prop}
Suppose the kernel function for the RKHS, $\mathcal{H}$ is isotropic. Then
\begin{equation}
	\frac{\delta^{\mathcal{H}} \bar{H}_F}{\delta \bar{f}} = \frac{1}{2} | v |^2 - \Delta_x^{-1} \int \bar{f} \mathsf{d} v \,.
\end{equation}
\end{prop}
\noindent \textit{Proof:} This follows from the fact that $\delta^{\mathcal{H}} \bar{F} / \delta \bar{f} = \mathcal{L} \delta F / \delta \bar{f}$, the previous lemma, and the fact that
\begin{equation}
	\bar{H}_F [\bar{f}] 
		= \frac{1}{2} \int \left( | v |^2 - \Delta_x^{-1} \int \bar{f}(x, v') \mathsf{d} v' \right) \mathcal{L}^{-1} \bar{f}(x, v) \mathsf{d} x \mathsf{d} v \,.
\end{equation} 
\qed

\noindent From this, one obtains the equations of motion implied by the filtered Vlasov Hamiltonian and Poisson bracket.

\begin{prop}
The continuous filtered Vlasov Hamiltonian structure yields the equation of motion
\begin{equation}
	\partial_t f = \left[ f, \frac{1}{2} | v |^2 - \Delta_x^{-1} \int \mathcal{L} f \mathsf{d} v \right] \,,
\end{equation}
where $f$ is the unfiltered distribution function. 
\end{prop}
\noindent \textit{Proof:} For any functional $\bar{F}$ of $\bar{f}$, one has
\begin{equation}
\begin{aligned}
	\dot{\bar{F}} = \{ \bar{F}, \bar{H} \}_{\mathcal{H}} 
	&= \left(\bar{f}, \left[\frac{\delta^\mathcal{H} \bar{F} }{\delta \bar{f}},  \frac{1}{2} | v |^2 - \Delta_x^{-1} \int \bar{f} \mathsf{d} v \right] \right)_{\mathcal{H}} \\
	&= \left(f, \left[ \frac{\delta F }{\delta f},  \frac{1}{2} | v |^2 - \Delta_x^{-1} \int \bar{f} \mathsf{d} v \right] \right)_{L^2} \\
	&= \left(\frac{\delta F }{\delta f}, \left[ f,  \frac{1}{2} | v |^2 - \Delta_x^{-1} \int \bar{f} \mathsf{d} v \right] \right)_{L^2} 
	= \dot{F} \,.
\end{aligned}
\end{equation}
The result follows. \qed

\noindent Hence, the Hamiltonian structure reformulated in the RKHS yields a Vlasov equation in which the charge distribution has been spatially smoothed prior to solving Poisson's equation. 

This continuous Hamiltonian structure yields an exact reduction which corresponds with the generic filtered particle method when representing the distribution function as a sum of spatially filtered particles. 
\begin{prop}
Let the filtered distribution function take the form $\bar{f} = \sum_a w_a k( | x_a - x | ) \delta(v - v_a)$ where $\int k \mathsf{d} x = 1$. Then the reduced particle Hamiltonian is given by
\begin{equation}
	H_p(X, V) = \frac{1}{2} \sum_{a=1}^{N_p} w_a | v_a |^2 - \frac{1}{2} \sum_{ \substack{a,b = 1 \\ a \neq b} }^{N_p} w_a w_b V ( | x_a - x_b | ) \,,
\end{equation}
where the potential is given by $V = \Delta_x^{-1} k$. 
\end{prop}
\noindent \textit{Proof:} That the kinetic energy becomes
\begin{equation}
	\frac{1}{2} \left( \bar{f}, | v |^2 \right)_{\mathcal{H}} = \frac{1}{2} \sum_{a=1}^{N_p} w_a | v_a |^2 
\end{equation}
is clear. On the other hand, the potential energy becomes
\begin{equation}
	- \frac{1}{2} \left( \Delta_x^{-1} \int f \mathsf{d} v, f \right)_{\mathcal{H}} 
	= - \frac{1}{2} \left( \Delta_x^{-1} \int f \mathsf{d} v , \mathcal{L}^{-1} f \right)_{L^2}
	= - \frac{1}{2} \sum_{a,b = 1}^{N_p} w_a w_b \Delta_x^{-1} \left. K_{x_a}(x) \right|_{x = x_b}
\end{equation}
since $\mathcal{L}^{-1} f = \sum_a w_a \delta(x - x_a) \delta(v - v_a)$. But $V(x_a - x_b) = \Delta_x^{-1} \left. K_{x_a}(x) \right|_{x = x_b}$ is a constant when $a = b$. Hence, these do not contribute to the dynamics and their contributions to the sum may safely be neglected. \qed

As shown in the previous section, letting $\bar{F}[\bar{f}] = F_p(Z)$, the Poisson bracket simply becomes
\begin{equation}
	\{\bar{F}, \bar{G}\}_{\mathcal{H}} 
	= \left( \bar{f}, \left[ \frac{\delta^\mathcal{H} \bar{F}}{\delta \bar{f}},  
		\frac{\delta^\mathcal{H} \bar{G}}{\delta \bar{f}} \right] \right)_{\mathcal{H}} 
	= \sum_a \frac{1}{w_a} \left[ \frac{\partial F_p }{\partial Z}, \frac{\partial G_p }{\partial Z} \right] 
	\eqcolon [ F_p, G_p ] \,.
\end{equation}
Hence, the equations of motion associated with this bracket and Hamiltonian are
\begin{equation}
\begin{cases}
	\dot{x}_a = v_a \\
	\dot{v}_a = \nabla_{x_a} \phi (x_a) \,,
\end{cases}
\quad \forall a \in \{1, \hdots, N_p \} \,,
\end{equation}
where $ - \Delta_{x_a} \phi (x_a) = \sum_{b \neq a} w_b k( | x_a - x_b | )$. This is identical to the generic filtered particle method presented at the beginning of this section. 

\begin{ex}
Consider the following example in $(x,v) \in \mathbb{R} \times \mathbb{R}$ where the smoothing kernel is the Laplace-type distribution function discussed in a previous example. The filtered distribution function is then
\begin{equation}
	\bar{f}(x,v) = \sum_{a=1}^{N_p} \frac{w_a}{2 \alpha} \exp \left( - \frac{| x - x_a |}{\alpha} \right) \delta(v - v_a) \,.
\end{equation}
it follows that
\begin{equation}
	f(x,v) = (1 - \alpha^2 \partial_x^2) \bar{f}(x,v) = \sum_{a=1}^{N_p} w_a \delta( x - x_a) \delta(v - v_a).
\end{equation}
It may be shown that, at least in a distributional sense,
\begin{equation}
	\partial_x^2 \left[ \frac{1}{2} | x - x_a | + \frac{\alpha}{2} \exp \left( - \frac{ | x - x_a | }{\alpha} \right) \right]
	= \frac{1}{2 \alpha} \exp \left( - \frac{ | x - x_a | }{\alpha} \right) \,.
\end{equation}
Therefore, the particle Hamiltonian is given by
\begin{equation}
	H_p(Z) = \frac{1}{2}  \sum_{a=1}^{N_p} w_a \left( v_a^2 + \sum_{ \substack{b=1 \\ b \neq a} }^{N_p} w_b \left[ \frac{| x_a - x_b |}{2} 
		+ \frac{\alpha}{2} \exp \left( - \frac{ | x_a - x_b | }{\alpha} \right) \right] \right) \,.
\end{equation}
The evolution equation is given by
\begin{equation}
	\dot{F}(Z) = \sum_{a=1}^{N_p} \frac{1}{w_a} \left[ \frac{\partial F}{\partial Z}, \frac{\partial H_p}{\partial Z} \right].
\end{equation}
Note, the potential energy function has a continuous sigmoidal shaped derivative:
\begin{equation}
	\frac{\partial H_p(Z)}{\partial z_a} = \left( v_a, \sum_{b=1}^{N_p} w_b \left( \bm{1}_{x \geq 0}(x_a - x_b) 
		- \frac{1}{2} \right) \left[1 - \exp \left( - \frac{|x_a - x_b|}{\alpha} \right) \right] \right)
\end{equation}
where
\begin{equation}
	\bm{1}_{x \geq 0}(x) =
	\begin{cases}
		0, & x < 0 \\
		1, & x > 1 \,.
	\end{cases}
\end{equation}
The particles continue to influence each other's trajectory even if they are infinitely far from each other. Moreover, the force exerted by each particle approaches a constant value as the separation between particles goes to infinity. This is due to the one-dimensional character of this problem. Recall, the electric field due to an infinite sheet of charge remains constant as you move farther from the charge sheet. Each particle is really like an infinite sheet of charge in this one-dimensional setting. The effect of the smoothing kernel is that the transition of the electric field from one side of the charge sheet to the other is now a smooth sigmoidal function rather than a discontinuous jump. 
\end{ex}

The prior example allows one to see how the method works when the inner product structure associated with the RKHS is known as in the case when $\mathcal{H} = H^1(\mathbb{R})$. However, a Laplace-distribution-like smoothing kernel is hardly the most typical kernel function or the most useful smoothing kernel to choose (its derivative is discontinuous at the origin and the kernel has infinite support). General smoothing kernels may be used which yield RKHS whose inner product structure might not be explicitly written, but nonetheless is known to exist. Hence, this framework may be applied to discern the underlying Hamiltonian structure associated with any smoothing kernel one might choose, e.g.\ B-spline kernel functions. 

\section{Application to general mean-field Hamiltonian theories}

The argument showing that smoothed particle discretizations of the Vlasov-Poisson equation are an exact reduction of a continuous Lie-Poisson system with a regularized Hamiltonian readily applies to other Hamiltonian theories which are generically called mean-field theories \cite{Morrison2003}. Let $Q \subset \mathbb{R}^{2d}$, where $d \in \mathbb{Z}^+$, and denote the finite dimensional canonical Poisson bracket by $[\cdot, \cdot ]: C^\infty(Q) \times C^\infty(Q) \to \mathbb{R}$. The mean-field Hamiltonian field theories take the form
\begin{equation}
    \{F, G\} 
    =
    \int \omega \left[ \frac{\delta F}{\delta \omega}, \frac{\delta G}{\delta \omega} \right] \mathsf{d} z
    \quad \text{and} \quad
    H[\omega] = 
    \int h_0(z) \omega(z) \mathsf{d} z
    +
    \frac{1}{2} \int h_1(z,z') \omega(z) \omega(z') \mathsf{d} z \mathsf{d} z' \,,
\end{equation}
where $h_0(z)$ is interpreted as the single particle energy and $h_1(z, z')$ is an interaction potential. This Hamiltonian and Poisson bracket yields the following equations of motion:
\begin{equation}
    \partial_t \omega 
    + \left[ h_0 + \int h_1(z, z') \omega(z') \mathsf{d} z', \omega \right] = 0 \,.
\end{equation}
In addition to the Vlasov-Poisson equation, Darwin's approximation \cite{10.1063/1.2799346}, the two-dimensional vorticity equation \cite{pjm82, osti_6351319}, and the Charney-Hasegawa-Mima equation \cite{10.1063/1.3194275, CHANDRE2014956} (which is structurally equivalent to the quasigeostrophic potential vorticity equation) may all be written as mean-field Hamiltonian systems. The conditions which allow these mean-field Hamiltonian models to admit a regularization which has a finite dimensional exact Hamiltonian reduction based on smoothed particles will is briefly discussed below.

One may rewrite the above Hamiltonian model as follows
\begin{equation}
    \{F, G\}
    =
    \left( \omega, 
    \left[ \frac{\delta F}{\delta \omega}, \frac{\delta G}{\delta \omega} \right] \right)_{L^2}
    \quad \text{and} \quad
    H[\omega]
    =
    \left( \omega, h_0 + V \omega \right)_{L^2}
\end{equation}
where the interaction operator is defined
\begin{equation}
    V \omega
    = \int h_1(z,z') \omega(z') \mathsf{d} z' \,.
\end{equation}
If $\mathcal{L} \omega = k * \omega$ where $k$ is the kernel function, then the filtered Hamiltonian is defined
\begin{equation}
    H_F[\omega]
    =
    \left( \omega, h_0 + V \mathcal{L} \omega \right)_{L^2} \,.
\end{equation}
As in the Vlasov-Poisson equation, one requires that $V$ and $\mathcal{L}$ commute. If this is so, then the above Hamiltonian structure yields the equations of motion
\begin{equation}
    \partial_t \omega 
    + \left[ h_0 + \int h_1(z, z') \mathcal{L} \omega(z') \mathsf{d} z', \omega \right] = 0 \,.
\end{equation}
This is identical to the equations of motion in the unsmoothed case except that the convolution operator is applied prior to computing the interaction potential. As in the case of the Vlasov-Poisson equation, one may equivalently interpret the regularized Hamiltonian model in terms of a RKHS, $(\mathcal{H}, (\cdot, \cdot)_{\mathcal{H}})$, associated with the kernel function, $k$. Letting $\bar{\omega} = \mathcal{L} \omega$, the regularized theory equivalently may be written
\begin{equation}
    \{ \bar{F}, \bar{G}\}_{\mathcal{H}}
    =
    \left( \bar{\omega}, 
    \left[ \frac{\delta^{\mathcal{H}} \bar{F}}{\delta \bar{\omega}}, \frac{\delta^{\mathcal{H}} \bar{G}}{\delta \bar{\omega}} \right] \right)_{\mathcal{H}}
    \quad \text{and} \quad
    \bar{H}_F[\bar{\omega}]
    =
    \left( \bar{\omega}, h_0 + V \bar{\omega} \right)_{\mathcal{H}} \,.
\end{equation}
The two formulations yield equivalent equations of motion. 

By letting 
\begin{equation}
    \bar{\omega}(z) = \sum_a w_a k(z - z_a) \,,
    \quad \text{or} \quad
    \omega(z) = \sum_a w_a \delta(z - z_a) \,,
\end{equation}
one obtains a finite dimensional Hamiltonian system prescribing the evolution of the particle centers $Z = (z_1, \hdots, z_{N_p})$ with Hamiltonian
\begin{equation}
    H[Z] = \sum_a w_a h_0(z_a)
    + \sum_{a,b} w_a w_b \int h_1(z_a, z) k(z - z_b) \mathsf{d} z \,,
\end{equation}
and Poisson bracket
\begin{equation}
    \{ F_p, G_p \} = \sum_a \frac{1}{w_a} \left[ \frac{\partial F_p }{\partial Z}, \frac{\partial G_p }{\partial Z} \right] \,.
\end{equation}

In the case of two-dimensional vorticity, $Q = \mathbb{R}^2$, $h_0(x,y) = 0$, and the interaction potential is given by
\begin{equation}
    V \omega = \Delta^{-1} \omega \,. 
\end{equation}
Hence, a result very similar to Lemma \ref{lemma:self_adjoint} applies to show that any isotropic, symmetric, positive definite kernel function might be used to obtain a smoothed point-vortex model. The study of point vortex dynamics has a long history, but a smoothed theory of point vortex dynamics analogous to a smoothed PIC method appears to be a novel method suggested by this paper. Likewise, each of the mean-field Hamiltonian theories mentioned at the start of this section also yield a smoothed particle discretization. 

\section{A filtered Vlasov-Maxwell system and its Hamiltonian structure}
It is also possible to derive a filtered PIC discretization of the Vlasov-Maxwell system which is an exact reduction of a continuous Hamiltonian system. The approach differs from that taken for general mean-field theories because the Hamiltonian is not quadratic in the distribution function. Whereas mean-field theories are filtered by modifying the Hamiltonian, the Vlasov-Maxwell equation is filtered by modifying the Poisson bracket. One shall see that this change amounts to a coordinate change. 

The Vlasov-Maxwell equation was shown to be a Hamiltonian system \cite{MORRISON1980383, pjm82, MARSDEN1982394}, and Hamiltonian structure-preserving PIC discretizations have been developed in the past decade, e.g. \cite{10.1063/1.4742985, kraus_kormann_morrison_sonnendrücker_2017}. Likewise, when inserting smoothing, it is desirable to ascertain the right way of doing so such that the salient geometric features of the model are retained. To this end, it is desirable to retain a variational principle in canonical coordinates. Smoothed structure-preserving PIC discretizations built directly from the variational principle are known \cite{variational_PIC_campos-pinto}. Rather than considering the variational principle directly, this work instead considers to the Hamiltonian formulation in canonical coordinates.

The Poisson bracket of the Vlasov-Maxwell system may be written as follows \cite{MORRISON1980383, pjm82, MARSDEN1982394}:
\begin{equation} \label{eq:vlasov-maxwell_PB}
    \begin{aligned}
        \{F, G\}
        &= 
        \int f \left[ \frac{\delta F}{\delta f}, \frac{\delta G}{\delta f} \right] \mathsf{d} x \mathsf{d} v
        + 
        \int \left( \frac{\delta F}{\delta E} \cdot \frac{\partial f}{\partial v} \frac{\delta G}{\delta f} - \frac{\delta G}{\delta E} \cdot \frac{\partial f}{\partial v} \frac{\delta F}{\delta f} \right) \mathsf{d} x \mathsf{d} v \\
        &+ \int f B \cdot \left( \frac{\partial}{\partial v} \frac{\delta F}{\delta f} \times \frac{\partial}{\partial v} \frac{\delta G}{\delta f} \right) \mathsf{d} x \mathsf{d} v
        + \int \left( \frac{\delta F}{\delta E} \cdot \nabla \times \frac{\delta G}{\delta B} - \frac{\delta G}{\delta E} \cdot \nabla \times \frac{\delta F}{\delta B} \right) \mathsf{d} x \,,
    \end{aligned}
\end{equation}
and its Hamiltonian is written
\begin{equation} \label{eq:ham_vlasov-maxwell}
    H[f, E, B] =
    \frac{1}{2} \int | v |^2 f(x, v) \mathsf{d} x \mathsf{d} v
    + \frac{1}{2} \int ( | E |^2 + | B |^2 ) \mathsf{d} x \,.
\end{equation}
As in the rest of this article, all constants have been set to unity. 

The Vlasov-Maxwell equation may also be written in canonical coordinates such that it is the direct sum of a Poisson bracket which generates the Vlasov equation and a Poisson bracket which generates Maxwell's equations:
\begin{equation}
    \{F, G\} 
    =
    \int f_{can} \left[ \frac{\delta F}{\delta f_{can}}, \frac{\delta G}{\delta f_{can}} \right] \mathsf{d} x \mathsf{d} p
    -
    \int \left( \frac{\delta F}{\delta E} \cdot \frac{\delta G}{\delta A} - \frac{\delta G}{\delta E} \cdot \frac{\delta F}{\delta A} \right) \mathsf{d} x \,,
\end{equation}
where $A$ is the vector potential, the canonical momentum is $p = v + A$, and the distribution function in canonical coordinates related to the usual phase space distribution by
\begin{equation}
    f(x,v) = f_{can}(x, v + A(x)) \,.
\end{equation}
The Hamiltonian in canonical coordinates is
\begin{equation}
    H[f_{can}, A, E] = 
    \frac{1}{2} \int | p - A(x) |^2 f_{can}(x, p) \mathsf{d} x \mathsf{d} p
    + \frac{1}{2} \int ( | E |^2 + | \nabla \times A |^2 ) \mathsf{d} x \,.
\end{equation}
In order to recover the noncanonical Vlasov-Maxwell bracket from the canonical bracket, one makes the change of coordinates
\begin{equation}
    F[f_{can}, A, E] = \bar{F}[f, B, E] \,,
\end{equation}
where $B = \nabla \times A$, which was shown to be a Hamiltonian reduction \cite{MARSDEN1982394}.  

Smoothing of the Vlasov-Maxwell equation should only appear in the coupling terms which give rise to the current in Maxwell's equations, and the Lorentz force in the Vlasov equation. It turns out that this may be accomplished by defining a smoothed canonical phase space distribution as follows:
\begin{equation}
    \bar{f}_{can}(x, v + \mathcal{L} A(x)) = f(x,v) \,.
\end{equation}
That is, one simply applies smoothing to the vector potential appearing in the the transformation from canonical to noncanonical coordinates. As usual, one assumes $\mathcal{L}$ is a symmetric positive definite kernel. 

Several lemmas are provided below which facilitate the change of coordinates. 
\begin{lemma}
Suppose 
\begin{equation}
    \bar{F}[\bar{f}_{can}, A, E] = F[f, B, E] \,,
\end{equation}
where $\bar{f}_{can}(x, v + \mathcal{L} A(x)) = f(x,v)$, and $\nabla \times A = B$. Then
\begin{equation}
    \frac{\delta \bar{F}}{\delta \bar{f}_{can}}
    =
    \frac{\delta F}{\delta f} \,,
    \quad
    \frac{\delta \bar{F}}{\delta A}
    =
    \nabla \times \frac{\delta F}{\delta B}
    +
    \mathcal{L} \left( \frac{\partial f}{\partial v}
    \frac{\delta F}{\delta f} \right) \,,
    \quad \text{and} \quad
    \frac{\delta \bar{F}}{\delta E}
    =
    \frac{\delta F}{\delta E} \,.
\end{equation}
\end{lemma}
\noindent \textit{Proof:} This is a consequence of the chain rule. To be explicit, one has that
\begin{equation}
    \begin{pmatrix}
        \dfrac{\delta f}{\delta \bar{f}_{can}} & 
            \dfrac{\delta f}{\delta A} &
                \dfrac{\delta f}{\delta E} \\
        \dfrac{\delta B}{\delta \bar{f}_{can}} & 
            \dfrac{\delta B}{\delta A} &
                \dfrac{\delta B}{\delta E} \\
        \dfrac{\delta E}{\delta \bar{f}_{can}} & 
            \dfrac{\delta E}{\delta A} &
                \dfrac{\delta E}{\delta E}
    \end{pmatrix}
    =
    \begin{pmatrix}
        I & \frac{\partial f}{\partial v} \cdot \mathcal{L} & 0 \\
        0 & \nabla \times & 0 \\
        0 & 0 & I
    \end{pmatrix} .
\end{equation}
Hence, one finds
\begin{multline}
    \int \frac{\delta \bar{F}}{\delta \bar{f}_{can}} \delta \bar{f}_{can} \mathsf{d} x \mathsf{d} p
        + \int \left( 
            \frac{\delta \bar{F}}{\delta A} \cdot \delta A 
            + \frac{ \delta \bar{F}}{\delta E} \cdot \delta E \right) \mathsf{d} x \\
    =
    \int \frac{\delta F}{\delta f} \delta \bar{f}_{can} \mathsf{d} x \mathsf{d} p
        + \int \left[ \left( \nabla \times \frac{\delta F}{\delta B} 
            + \mathcal{L} \left( \frac{\delta F}{\delta f} \frac{\partial f}{\partial v} \right) 
                \right) \cdot \delta A
            + \frac{\delta F}{\delta E} \cdot \delta E \right] \mathsf{d} x \,.
\end{multline}
The result follows. \qed

\begin{lemma}
Suppose $\mathcal{L} g = k * g$. Moreover, assume $g, k$ are differentiable. Then
\begin{equation}
    \frac{\partial}{\partial x} (\mathcal{L} g) = \frac{\partial k}{\partial x} * g = k * \frac{\partial g}{\partial x} \,.
\end{equation}
\end{lemma}

\noindent With these lemmas, one may change coordinates in the Vlasov bracket.

\begin{lemma}
    Let $B = \nabla \times A$, and $g(x,v) = \bar{g}_{can}(x, v + \mathcal{L} A)$, and $h(x,v) = \bar{h}_{can}(x, v + \mathcal{L} A)$. Then
    \begin{equation}
        [\bar{g}_{can}, \bar{h}_{can} ]_{xp}
        =
        [g, h ]_{xv} + \mathcal{L} B \cdot \left( \frac{\partial g}{\partial v} \times \frac{\partial h}{\partial v} \right) \,,
    \end{equation}
    where $[\cdot, \cdot]_{xp}$ and $[\cdot, \cdot]_{xv}$ are respectively the canonical brackets with the coordinates $(x,p)$ and $(x,v)$. 
\end{lemma}

\noindent \textit{Proof:} Using the fact that $v = p - \mathcal{L} A(x)$ and the previous lemma, one has that
\begin{equation}
    \frac{\partial \bar{g}_{can}}{\partial x}
    =
    \frac{\partial g}{\partial x} 
        - \mathcal{L} \left( \frac{\partial A}{\partial x} \right) \frac{\partial g}{\partial v}
    \quad \text{and} \quad
    \frac{\partial \bar{g}_{can}}{\partial p} 
    =
    \frac{\partial g}{\partial v} \,.
\end{equation}
Plugging this into the canonical bracket, one obtains the result. \qed

\begin{theorem}
Suppose one lets
\begin{equation}
    \bar{F}[\bar{f}_{can}, A, E] = F[f, B, E] \,,
    \quad \text{and} \quad
    \bar{G}[\bar{f}_{can}, A, E] = G[f, B, E] \,,
\end{equation}
where $\bar{f}_{can}(x, v + \mathcal{L} A(x)) = f(x,v)$, and $\nabla \times A = B$. Then under this change of coordinates, the canonical bracket, 
\begin{equation}
    \{ \bar{F}, \bar{G} \}_{can} =
    \int \bar{f}_{can} \left[ \frac{\delta \bar{F}}{\delta \bar{f}_{can}}, \frac{\delta \bar{G}}{\delta \bar{f}_{can}} \right]_{xp} \mathsf{d} x \mathsf{d} p
    -
    4 \pi \int \left( \frac{\delta \bar{F}}{\delta E} \cdot \frac{\delta \bar{G}}{\delta A} - \frac{\delta \bar{G}}{\delta E} \cdot \frac{\delta \bar{F}}{\delta A} \right) \mathsf{d} x \,,
\end{equation}
becomes
\begin{equation} \label{eq:filtered_vlasov_maxwell_PB}
    \begin{aligned}
        \{F, G\}
        &= 
        \int f \left[ \frac{\delta F}{\delta f}, \frac{\delta G}{\delta f} \right]_{xv} \mathsf{d} x \mathsf{d} v
        + 
        \int \left( \frac{\delta F}{\delta E} \cdot \mathcal{L} \left( \frac{\partial f}{\partial v} \frac{\delta G}{\delta f} \right) - \frac{\delta G}{\delta E} \cdot \mathcal{L} \left( \frac{\partial f}{\partial v} \frac{\delta F}{\delta f} \right) \right) \mathsf{d} x \mathsf{d} v \\
        &+ \int f \mathcal{L} B \cdot \left( \frac{\partial}{\partial v} \frac{\delta F}{\delta f} \times \frac{\partial}{\partial v} \frac{\delta G}{\delta f} \right) \mathsf{d} x \mathsf{d} v
        + \int \left( \frac{\delta F}{\delta E} \cdot \nabla \times \frac{\delta G}{\delta B} - \frac{\delta G}{\delta E} \cdot \nabla \times \frac{\delta F}{\delta B} \right) \mathsf{d} x \,.
    \end{aligned}
\end{equation}
\end{theorem}

\noindent \textit{Proof:} This is an immediately consequence of the preceding lemmas. \qed

This manner of inserting spatial smoothing into the Vlasov-Maxwell system preserves the Hamiltonian structure. Because the canonical system was Hamiltonian, and one has simply changed coordinates, the noncanonical system is also Hamiltonian. As desired, the only changes made to the bracket are in the coupling terms between the particle and electromagnetic subsystems. 

The derivatives of the Hamiltonian are simple to compute:
\begin{equation}
    \frac{\delta H}{\delta f} = \frac{1}{2} | v |^2 \,,
    \quad
    \frac{\delta H}{\delta E} = E \,,
    \quad \text{and} \quad
    \frac{\delta H}{\delta B} = B \,.
\end{equation}
The equations of motion are obtained by noting that, for any functional of the fields, $F[f, B, E]$, one has $\dot{F} = \{F, H \}$. If one assumes that the kernel smoother is the identity operator in the velocity coordinate and only applies smoothing in space, one obtain the filtered Vlasov-Maxwell equations:
\begin{equation}
    \begin{aligned}
        \partial_t f &= - v \cdot \frac{\partial f}{\partial x} - \left( \mathcal{L} E + v \times \mathcal{L} B \right) \cdot \frac{\partial f}{\partial v} \\
        \partial_t E &= \nabla \times B - \mathcal{L} \int v f \mathsf{d} v \\
        \partial_t B &= - \nabla \times E \,.
    \end{aligned}
\end{equation}
It was necessary to assume the smoothing operator only applied smoothing in the spatial coordinate in order to integrate by parts in velocity space when obtaining the above evolution equations from the Hamiltonian formulation. As desired, filtering only occurs in those terms which couple the particle and field subsystems, namely the Lorentz force and the current. 

\section{Filtered Vlasov-Maxwell: a second interpretation}
There is a second manner by which one might obtain the filtered Vlasov-Maxwell Hamiltonian structure by directly mollifying the electromagnetic fields. To do so, one begins with the usual Vlasov-Maxwell Poisson bracket given in equation \eqref{eq:vlasov-maxwell_PB}, and then make the change of coordinates
\begin{equation}
    B \mapsto \mathcal{L} B
    \quad \text{and} \quad
    E \mapsto \mathcal{L}^{-1} E \,.
\end{equation}
It seems strange to apply smoothing to $B$ while applying anti-smoothing to $E$. However, one finds that this change of coordinates recovers exactly the filtered Vlasov-Maxwell Poisson bracket given in equation \eqref{eq:filtered_vlasov_maxwell_PB}. One does not apply this change of coordinates in the Hamiltonian however so that it remains as in equation \eqref{eq:ham_vlasov-maxwell}. Therefore, while the bracket is obtained via a simple change of coordinates, the Hamiltonian is kept formally the same. To build intuition for why it is appropriate to apply an anti-smoothing operator to the electric field, consider Gauss's law:
\begin{equation}
    \nabla \cdot E = \rho \mapsto \nabla \cdot \mathcal{L}^{-1} E = \rho 
    \quad \text{or} \quad
    \nabla \cdot E = \mathcal{L} \rho \,.
\end{equation}
This anti-smoothing operation on the electric field yields a smoothed charge density in Gauss's law. 

\section{A filtered GEMPIC discretization of the Vlasov-Maxwell system}
Using the filtered Vlasov-Maxwell system derived in the previous section, it is possible to derive a structure-preserving PIC discretization of this system in which the charge and current sources in Maxwell's equations have been smoothed. For this purpose, the approach taken in the Geometric Electromagnetic PIC (GEMPIC) method \cite{kraus_kormann_morrison_sonnendrücker_2017} is closely followed since only a few pieces of the algorithm need to change to accommodate filtering. For brevity, this method is called filtered-GEMPIC. Only enough detail is provided to show how the filtered-GEMPIC method modifies the GEMPIC method in a few places. 

As in GEMPIC, a finite element exterior calculus (FEEC) discretization \cite{feec_1, arnold_falk_winther_2006, BUFFA20101143} is used for the fields because of their utility in preserving the Hamiltonian structure of electrodynamics (specifically, this method conserves Gauss's laws), and a particle-based discretization is used in space. The properties of a FEEC discretization may be summarized by the following commuting diagram of vector spaces
\begin{equation} \label{eq:disc_derham_diagram}
    \begin{tikzcd}
    V^0 = H^1 \arrow{r}{\nabla} \arrow{d}{\Pi_0} & 
    V^{1} = H(\text{curl}) \arrow{r}{\nabla \times} \arrow{d}{\Pi_1} & 
    V^{2} = H(\text{div}) \arrow{r}{\nabla \cdot} \arrow{d}{\Pi_2} &
    V^3 = L^2  \arrow{d}{\Pi_3}\\
    V^0_h \arrow{r}{\nabla} & 
    V^{1}_h \arrow{r}{\nabla \times} & 
    V^{2}_h \arrow{r}{\nabla \cdot} &
    V^3_h \,,
    \end{tikzcd}
\end{equation}
where $V^\ell$ are the infinite dimensional vector spaces for the continuous fields, and $V^\ell_h$ are finite element subspaces. The projection operators $\Pi_\ell: V^\ell \to V^\ell_h$ have the properties $\nabla \Pi_0 \phi = \Pi_1 \nabla \phi$, $\nabla \times \Pi_1 E = \Pi_2 \nabla \times E$, and $\nabla \cdot \Pi_2 B = \Pi_3 \nabla \cdot B$, i.e. the diagram commutes. This feature of FEEC methods is essential for the structure-preserving discretization of the electromagnetic subsystem of the Vlasov-Maxwell system. The commutativity ensures that vector calculus relations $\nabla \times \nabla \equiv 0$ and $\nabla \cdot \nabla \times \equiv 0$ are exactly preserved in the finite element subspaces. 

The basis functions of $V^\ell_h$ are denoted by $\{ \Lambda_i^\ell \}_{i=1}^{N_\ell}$ where $\dim( V_h^\ell) = N_\ell$. The finite element mass matrices are defined as follows:
\begin{equation}
    (\mathbb{M}_\ell)_{ij} = ( \Lambda_i^\ell, \Lambda_j^\ell )_{L^2} \,.
\end{equation}
These matrices represent $L^2$ inner products at the coefficient level. The coefficients of a field in $V_h^\ell$ are denoted with boldface, $\bm{\mathsf{A}} = (\mathsf{A}_1, \hdots, \mathsf{A}_{N_\ell})$, so that if $A_h \in V_h^\ell$, it may be written
\begin{equation}
    A_h(x) = \sum_{i=1}^{N_\ell} \mathsf{A}_i \Lambda_i^\ell(x) \,.
\end{equation}
Given a field in the finite element space, one recovers the coefficient vector using the degrees of freedom operators which are denoted by $\sigma^\ell$. These are defined such that
\begin{equation}
    \sigma^\ell_i ( \Lambda^\ell_j ) = \delta_{ij}
    \implies
    \sigma^\ell_i(A_h) = \mathsf{A}_i \,.
\end{equation}
The precise form of these operators depends on the interpolating basis used. See \cite{kraus_kormann_morrison_sonnendrücker_2017} for the details of their construction in the context of a B-spline FEEC method. Finally, the matrices defining the discrete differential operators are defined as follows:
\begin{equation}
    \mathbb{G}_{ij} = \sigma^1_i( \nabla \Lambda_j^0 ) \,,
    \quad
    \mathbb{C}_{ij} = \sigma^2_i( \nabla \times \Lambda_j^1 ) \,,
    \quad \text{and} \quad 
    \mathbb{D}_{ij} = \sigma^3_i( \nabla \cdot \Lambda_j^2 ) \,.
\end{equation}
Finally, the particle position, velocity, and phase space coordinates are denoted by $\bm{X} = (x_1, \hdots, x_{N_p})$, $\bm{V} = (v_1, \hdots, v_{N_p})$, and $\bm{Z} = (z_1, \hdots, z_{N_p})$ respectively. 

To discretize the model, one simply posits that the fields and distribution function take a particular form which depends only on finite dimensional coefficient vectors. Take $E_h = \Pi_1 E \in V^1_h$ and $B_h = \Pi_2 B \in V^2_h$, and let the phase space distribution be
\begin{equation}
    f_h(x,v) = \sum_a w_a \delta(x - x_a) \delta(v - v_a) \,.
\end{equation}
The discretization proceeds by defining the finite dimensional functions $\mathsf{F}$ and $\mathsf{G}$ to be functions of the coefficients of these finite dimensional representations:
\begin{equation}
    \mathsf{F}(\bm{X}, \bm{V}, \bm{\mathsf{E}}, \bm{\mathsf{B}})
    =
    F[f_h, E_h, B_h] \,,
\end{equation}
and similarly for $\mathsf{G}$. Plugging in these finite dimensional representations of the fields and following the approach developed in \cite{kraus_kormann_morrison_sonnendrücker_2017}, one finds the following finite dimensional Poisson bracket: 
\begin{equation}
    \begin{aligned}
        \{ F, G \}
        &= 
        \sum_a \frac{1}{w_a} \left[ \frac{\partial F}{\partial z_a}, \frac{\partial G}{\partial z_a} \right]
        + 
        \sum_a \frac{1}{w_a} \left[ 
        \mathbb{M}_1^{-1} \frac{\partial G}{\partial \bm{\mathsf{E}}} \cdot \bar{\bm{\Lambda}}^1(x_a) \frac{\partial F}{\partial v_a} -
        \mathbb{M}_1^{-1} \frac{\partial F}{\partial \bm{\mathsf{E}}} \cdot \bar{\bm{\Lambda}}^1(x_a) \frac{\partial G}{\partial v_a}
        \right] \\
        &+ \sum_a \frac{1}{w_a} \bar{B}_h(x_a) \cdot \left( \frac{\partial F}{\partial v_a} \times \frac{\partial G}{\partial v_a} \right)
        + \frac{\partial F}{\partial \bm{\mathsf{E}}} \cdot \mathbb{M}_1^{-1} \mathbb{C}^T \frac{\partial G}{\partial \bm{\mathsf{B}}} 
        -
        \frac{\partial G}{\partial \bm{\mathsf{E}}} \cdot \mathbb{M}_1^{-1} \mathbb{C}^T \frac{\partial F}{\partial \bm{\mathsf{B}}} 
        \,,
    \end{aligned}
\end{equation}
where the smoothed magnetic field is given by
\begin{equation} \label{eq:filtered_B}
    \bar{B}_h(x) = \int B_h(x') k(x - x') \mathsf{d} x'
    \quad \text{and} \quad
    B_h(x) = \sum_{i=1}^{N_2} \mathsf{B}_i \Lambda_i^2(x) \,,
\end{equation}
and the smoothed $1$-form basis functions are given by
\begin{equation} \label{eq:filtered_1-form}
    \bar{\Lambda}^1_i(x) = \int \Lambda^1_i(x') k(x - x') \mathsf{d} x'
    \quad \text{and} \quad
    \bar{\bm{\Lambda}}^1(x) =
    (\bar{\Lambda}^1_1(x), \hdots, \bar{\Lambda}^1_{N_1}(x) ) \,.
\end{equation}
The Hamiltonian is simply
\begin{equation}
    \mathsf{H}(\bm{X}, \bm{V}, \bm{\mathsf{E}}, \bm{\mathsf{B}})
    =
    \frac{1}{2} \sum_a w_a v_a^2
    +
    \bm{\mathsf{E}}^T \mathbb{M}_1 \bm{\mathsf{E}} + 
    \bm{\mathsf{B}}^T \mathbb{M}_2 \bm{\mathsf{B}} \,.
\end{equation}
Using the fact that for any any observable, $\mathsf{F}(\bm{X}, \bm{V}, \bm{\mathsf{E}}, \bm{\mathsf{B}})$, its evolution is computed as $\dot{\mathsf{F}} = \{\mathsf{F}, \mathsf{H} \}$, one obtains the equations of motion:
\begin{equation}
    \begin{aligned}
        \dot{x_a} &= v_a \\
        \dot{v_a} &= \frac{1}{w_a} \left( \bar{E}_h(x_a) + v_a \times \bar{B}_h(x_a) \right) \\
        \dot{\bm{\mathsf{E}}} &= \mathbb{M}_1^{-1} \left( \mathbb{C}^T \mathbb{M}_2 \bm{\mathsf{B}} - \sum_a \bar{\bm{\Lambda}}^1(x_a) v_a \right) \\
        \dot{\bm{\mathsf{B}}} &= - \mathbb{C} \bm{\mathsf{E}} \,,
    \end{aligned}
\end{equation}
where the filtered electric field is defined to be
\begin{equation}
    \bar{E}_h(x) = \int E_h(x') k(x - x') \mathsf{d} x'
    \quad \text{where} \quad
    E_h(x) = \sum_{i=1}^{N_1} \mathsf{E}_i \Lambda_i^1(x) \,,
\end{equation}
and the filtered magnetic field and $1$-form shape functions were previously defined in equations \eqref{eq:filtered_B} and \eqref{eq:filtered_1-form} respectively. 

If one eliminates the smoothing kernels, GEMPIC \cite{kraus_kormann_morrison_sonnendrücker_2017} is recovered. Moreover, the method ends up being identical to that found in \cite{variational_PIC_campos-pinto} which was obtained by inserting smoothing into a discretized Vlasov-Maxwell variational principle. It should not be surprising that the two approaches obtained the same PIC method as the variational principle and the Hamiltonian formulation are intimately related. Therefore, while filtered Hamiltonian structure-preserving PIC methods for the Vlasov-Maxwell system already exist in the literature, this paper elucidates the continuous Hamiltonian structure for which the PIC methods are exact Hamiltonian reductions. This provides an analytical tool for studying the impacts of smoothing on the model at the continuum level. Moreover, the results of this section suggest that, among the various ways of adding smoothing to an electromagnetic PIC method, the manner described in this section and in \cite{variational_PIC_campos-pinto} should be the preferred approach if one wishes the smoothed electromagnetic PIC method to be structure-preserving. 

\section{Conclusion}

This article provides a means of interpreting the Hamiltonian structure of smoothed Vlasov-Poisson PIC methods as a Lie-Poisson system in which the duality structure is taken to be the inner product of the RKHS associated with the smoothing kernel rather than the usual $L^2$ inner product. While the results are ultimately equivalent to the standard theory based on the $L^2$ inner product structure (because the smoothing convolution operation is designed to be invertible), there is utility in presenting this alternative perspective in which the distribution function is taken from a RKHS and expressed as a sum of kernel functions. Moreover, a small modification of the Hamiltonian after changing coordinates into the filtered representation yields a regularized Hamiltonian model in which the phase space distribution has been smoothed prior to solving Poisson's equation. Any symmetric, positive definite, isotropic filtering kernel which approximates the Dirac delta is compatible with the theory built herein. The theory may prove beneficial in leveraging the theory of RKHS, which has substantial utility in probability and statistics, for finding optimal strategies 
for estimating moments of the phase space distribution. It is possible to interpret filtering in a PIC method in terms of kernel density estimation \cite{10.1063/1.5038039}, a subject which in turn has profited from the theory of RKHS, e.g. for efficient density estimation in high dimensions \cite{doi:10.1137/22M147476X}. Moreover, kernel density estimation may be thought of as evaluation of a sample mean in the RKHS associated with the smoothing kernel \cite{JMLR:v13:kim12b}. The framework developed here, which provides a new interpretation of existing filtered PIC methods as arising from a Lie-Poisson Hamiltonian structure built from the inner product structure of a particular RKHS, might be beneficial in analyzing and designing structure-preserving filtered PIC methods because it makes explicit an alternative interpretation of the model in terms of a new RKHS inner product structure. In particular, the generality of this perspective may aid in the design of data-driven approaches to selecting optimal kernel functions. It was also shown that, generically, filtered PIC methods do not approximate the Vlasov-Poisson equation, but a smoothed version of it. It may be helpful to apply the Hamiltonian spectral theory previously developed for the Vlasov-Poisson system \cite{doi:10.1080/00411450008205881} to this regularized system in order to better understand how the stability theory of this smoothed system corresponds with the true system, and, by extension, how well one should expect filtered PIC methods to approximate the continuous Vlasov-Poisson system. Additionally, it was briefly shown how general mean-field Hamiltonian systems \cite{Morrison2003} are likewise amenable to the treatment applied in detail to the Vlasov-Poisson equation in this paper. This suggests the possibility of discretizations in the style of a smoothed PIC method for each of these systems. 

The Vlasov-Maxwell equation was shown to likewise admit a filtered version of its Hamiltonian structure although, unlike the Vlasov-Poisson equation, it is the Poisson bracket and not the Hamiltonian which is modified to facilitate this filtering. The connection with RKHS is less meaningful here than in the case of mean-field Hamiltonian models because one regularizes the electromagnetic fields with a smoothing kernel, not the phase space distribution itself. Therefore, the interpretation of the phase space distribution as a pointwise evaluation operator in an appropriate RKHS was not helpful. It was shown that the filtered Vlasov-Maxwell Hamiltonian structure may written in canonical coordiantes thus verifying that this is a valid Hamiltonian system. This model was then used to derive a filtered PIC discretization based on GEMPIC \cite{kraus_kormann_morrison_sonnendrücker_2017} which is a Hamiltonian structure preserving discretization of the continuous model. This PIC discretization was found to be identical to a previously found variational electromagnetic PIC method \cite{variational_PIC_campos-pinto} thus providing a new means of analyzing this computational model. 

As a final note, the manner of incorporating smoothed particles into the Hamiltonian structure of the Vlasov equation and related models presented herein suggests the possibility of incorporating smoothing kernels in particle models at a more fundamental level. The exact dynamics of a system of $N$ particles interacting through a force field is given by the Liouville equations, which describe the $6N$-dimensional particle distribution on phase space, or the Klimontovich equation, which describes the evolution of the characteristics of the particles. These two approaches describe the dynamics of a plasma in complete detail via slightly different perspectives. The Vlasov equation is obtained as a truncation of the BBGKY hierarchy, which is itself derived from the Liouville equation. The Hamiltonian structure of Liouville's equation is inherited by the BBGKY hierarchy, as demonstrated in \cite{bbgky84}, which is then inherited by the Vlasov equation, see \cite{pjm82, MORRISON1980383, MARSDEN1982394}. The Klimontovich equation describes the evolution of particles as delta-measures, but the methods described herein might be used to describe smoothed particle dynamics for the $N$-body model in a Hamiltonian context. In this manner, an alternative BBGKY hierarchy based on smoothed particles might be obtained which may yield a subtly different smoothed Vlasov equation when truncated. The consequences of replacing delta-measure particles with an delta-sequence family of smoothed particles in the Liouville equation and whether this small-scale regularization substantially impacts subsidiary models, e.g.\ the BBGKY hierarchy and Vlasov equation, is worth investigating in a future work.  

\section*{Acknowledgements}
The authors gratefully acknowledge the support of U.S. Dept. of Energy Contract \# DE-FG02-04ER54742, NSF Graduate Research Fellowship \# DGE-1610403.

\bibliography{references}{}
\bibliographystyle{plain}

\end{document}